\documentclass[reqno,11pt]{amsart}
\usepackage{amsmath,amssymb,tabularx,setspace,color,enumerate}
\usepackage{amsmath,amssymb,tabularx,setspace,color,graphicx,multirow,booktabs,tabularx,setspace,color,graphicx}
\usepackage[margin=3.5cm]{geometry}

\newtheorem{corollary}{Corollary}
\newtheorem{theorem}{Theorem}
\makeatletter
\renewcommand\section{\@startsection {section}{1}{\z@}%
                                   {-3.5ex \@plus -1ex \@minus -.2ex}%
                                   {2.3ex \@plus.2ex}%
                                   {\normalfont\large\bfseries}}
\usepackage{multirow}
\usepackage{array}
\usepackage{tabularx}
\usepackage{geometry}
\usepackage{amsmath}
\usepackage{setspace}
\usepackage{amsthm}
\usepackage{graphicx}
\usepackage{enumerate}
\begin{document}

\title[]{A new goodness of fit test for   uniform distribution with censored observations}
\author[]{S\lowercase{udheesh} K K\lowercase{attumannil}$^{\lowercase{a}}$ \lowercase{and}   S\lowercase{reedevi} E P$^{\lowercase{b},\dag}$ \\
$^{\lowercase{a}}$I\lowercase{ndian} S\lowercase{tatistical} I\lowercase{nstitute},
  C\lowercase{hennai}, I\lowercase{ndia,}\\
  $^{\lowercase{b}}$SNGS C\lowercase{ollege}, P\lowercase{attambi,} 
 I\lowercase{ndia.}}
\thanks{Corresponding author email: sreedeviep@gmail.com }
\doublespace
\maketitle\vspace{-0.2in}
 \begin{abstract}
 Using fixed point characterization, we develop a new goodness of fit test for uniform distribution. We also discuss how the right censored observations can be incorporated in the proposed test procedure. We study the asymptotic properties of the proposed test statistics. A Monte Carlo simulation is carried out  to evaluate the finite sample performance of the tests. We illustrate the test procedures using real data sets.
\end{abstract}\vspace{-0.2in}
\section{Introduction}
Uniform distribution is a widely used statistical model in various fields of applied as well as theoretical statistics. Due to the range property of standard uniform distribution, many Monte Carlo simulation algorithms use a sample from uniform distribution to generate random samples from other distributions. 
In view of the probability integral transformation, the simple goodness of fit problem of testing that a sample is from a particular continuous distribution is equivalent to testing that the transformed sample is from a uniform distribution on the interval $(0,1)$. This motivates us  to develop  a goodness of fit test for uniform distribution.  

Tests for uniform distribution has great significance in all discipline due to the property that the distribution with maximum entropy is standard uniform. Various tests were developed in literature to test the hypothesis of uniformity. We refer interested readers to  Kimball (1947), Sherman (1950),  Quesenberry and Miller Jr. (1977), Hegazy and Green (1975), Young (1982), Cheng and Spiring (1987) and Frozini et al. (1987). Under the uniform distribution assumption, Asgharian et al. (2002) estimated survival of dementia patients and de Una-Alvarez (2004) estimated the distribution of unemployment spells of women.  In many situations, we deal with a uniform distribution with truncation and/or censoring. A test for goodness of fit for a uniform truncation model was developed by  Mandel and Betensky (2007). Recently, Cho et al. (2021) developed a test for uniformity of exchangeable random variables on the circle.  However, goodness of fit for  uniform distribution under right censored case is not developed yet. Using  Stein's type characterisation we develop  goodness of fit tests for uniform distribution with complete as well as censored data.

Stein's identity and its application has been studied in literature.  See Sudheesh (2009) and Sudheesh and Isha (2016) for a generalized Stein's identity and its application. Making use of Stein's type identity,  Ebner and Liebenberg (2020) obtain the following fixed point characterization  for beta distribution.\vspace{-0.1in}
\begin{theorem}Let $X$ be a random variable (r.v.) taking values in $[0, 1]$. Then $X \sim B(\alpha, \beta)$ for $\alpha,\beta> 0$ if and only if
$$(\alpha+\beta)E(XI(X>t))=\alpha E(I(X>t))+\frac{t^{\alpha}(1-t)^{\beta}}{B(\alpha, \beta)},\,\,0\le t\le 1,$$where $I(A)$ denotes the indicator function of a set $A$.
\end{theorem}\vspace{-0.2in}
\begin{corollary}
 Let $X$ be a r.v. taking values in $[0, 1]$. Then $X \sim U(0,1)$ if and only if
\begin{equation*}\vspace{-0.1in}
  2E(XI(X>t))=E(I(X>t))+{t(1-t)},\,\,0\le t\le 1.
\end{equation*}
\end{corollary}Using the fixed point characterization given in Corollary 1, we develop a goodness of fit test for uniform distribution. The rest of the paper is organised as follows. In Section 2, we develop a new non-parametric test for uniform distribution. We study the asymptotic properties of the test statistic. In Section 3,  we discuss how to incorporate right censored observations in the proposed methodology. The results of the Monte Carlo simulation study are reported in Section 4. In Section 5, the test procedures are illustrated with application on real data sets. Finally, in Section 6, we give the concluding remarks.
\vspace{-0.3in}
\section{Test statistic}
In this section, we develop a goodness of fit test for uniform distribution for complete data.
Based on a random sample $X_{1}, ...,X_{n}$  from $U(0,1)$, we are interested in  testing the null hypothesis
$$H_{0}: F\text{ has uniform distribution}.$$
against$$ H_1: F\text{ does not follow uniform distribution}.$$
 For this purpose,  we define a departure measure which discriminate between null and alternative hypothesis. Consider  $\Delta(F)$ given by
\begin{eqnarray*}\label{deltam}
\Delta(F)&=&\int_{0}^{1}\left(2E(XI(X>t))-E(I(X>t))-{t(1-t)}\right)dF(t).
\end{eqnarray*}In view of Corollary 1, $\Delta(F)$ is zero under  $H_0$ and non zero under $H_1$. Hence $\Delta(F)$  can be considered as a measure of departure  from the null hypothesis $H_0$ towards the alternative  hypothesis $H_1$.

Next  we  express $\Delta(F)$ in a simple form as
\begin{eqnarray}\label{delta}
\Delta(F)&=&\int_{0}^{1}\left(2E(XI(X>t))-E(I(X>t))-t(1-t)\right)dF(t)\nonumber\\
&=&2\int_{0}^{1}\int_{0}^{1}yI(y>t)dF(y)dF(t)-  E(X)-E(X(1-X))\nonumber\\
&=&2\int_{0}^{1}\int_{t}^{1}ydF(y)dF(t)- 2E(X)+E(X^2)\nonumber\\
&=&2\int_{0}^{1}y\int_{0}^{y}dF(t)dF(y)- 2E(X)+E(X^2)\nonumber\\
&=&\int_{0}^{1}2yF{(y)}dF(y)-2E(X)-E(X^2)\nonumber\\
&=&E\big(max(X_{1},X_{2})-2X+X^2\big).
\end{eqnarray}Consider a symmetric kernel $h_1(X_1, X_2)=\frac{1}{2}(2\max(X_{1},X_{2})-2X_{1}-2X_{2}+X_{1}^2+X_{2}^2)$, then  $E(h_1(X_1, X_2))=\Delta(F)$. Hence we propose a U-statistic based test statistic given  by
$$\widehat\Delta=\frac{2}{n(n-1)}\sum_{i=1}^{n}\sum_{j=1,j<i}^{n}h_1(X_i,X_j),$$ for testing uniform distribution.  The $\widehat\Delta$ is an unbiased and consistent estimator of $\Delta(F)$ (Lehmann, 1951). Let $X_{(i)},\,\,i=1,\ldots,n$ be the order statistics based on $n$ independently and identically observations from $U(0,1)$. Then we can represent $\widehat\Delta$ in a simple form as
\begin{equation*}\label{deltahat}
  \widehat\Delta=\frac{1}{n(n-1)}\sum_{i=1}^{n}\left(2(i-n)+(n-1)X_{(i)}\right)X_{(i)}.
\end{equation*}
We reject the null hypothesis $H_0$ against the alternative  for large value of $|\widehat{\Delta}|$. Next we find an asymptotic critical region of the test using normal approximation. The following result is useful in this direction.
\begin{theorem}
  As $n\rightarrow \infty$,  $\sqrt{n}(\widehat{\Delta}-\Delta)$ converges in distribution to normal with mean zero and variance $\sigma^2$, where $\sigma^2$ is given by
  \begin{small}
 \begin{equation}\label{var}
\sigma^{2}=Var\Big(2XF(X)+2\int_{X}^{1}ydF(y)-2X+X^2\Big).
\end{equation}
\end{small}
\end{theorem}
\noindent {\bf Proof:}
Asymptotic  normality of $ \widehat{\Delta}$ follows from the central limit theorem for U-statistics. The asymptotic variance is $4\sigma_1^2$ where $\sigma_{1}^2$ is given by (Lee, 1990)
\begin{equation}\label{var1}
\sigma_1^2= Var\left[E\left(h(X_{1},X_{2}|X_{1})\right)\right].
\end{equation}Consider
\begin{eqnarray*}\label{var11}
E[\max(X_1,X_{2})|X_1=x]&=&E[x I(X_{2}\le x)+X_2I(X_{2}>x)]\nonumber \\
&=&xP(X_2\le x)+\int_{0}^{1}yI(x<y)dF(y)\nonumber\\
&=&xF(x)+\int_{x}^{1}ydF(y).
\end{eqnarray*}Hence, from (\ref{var1}) we obtain the variance expression given in (\ref{var}) and the proof of the theorem follows.\\
Under the null hypothesis $H_0$, we know that $\Delta{(F)}=0$. Hence we have the following corollary.
\begin{corollary}
 Under $H_0$, as $n\rightarrow \infty$,  $\sqrt{n}\widehat{\Delta}$ converges in distribution to normal with mean zero and variance $\frac{1}{45}$.
\end{corollary}
\noindent {\bf Proof:} Under $H_0$, we have
\begin{eqnarray*}
\sigma^{2}&=&Var\Big(2XF(X)+2\int_{X}^{1}ydF(y)-2X+X^2\Big)\\
&=&Var\Big(2X^2+2\int_{X}^{1}ydy-2X+X^2\Big)\\
&=&4Var\Big(X(X-1)\Big)=\frac{1}{45}.
\end{eqnarray*}


A distribution free test for testing uniform distribution  can be constructed  using Corollary 1. We reject the null hypothesis $H_{0}$ against the alternative hypothesis $H_{1}$ at a significance level $\alpha$, if
\begin{equation*}
 { \sqrt{45n} |\widehat{\Delta}| }>Z_{\alpha/2},
  \end{equation*}
where $Z_{\alpha}$ is the upper $\alpha$-percentile point of the standard normal distribution.  The performance of the test is evaluated through Monte Carlo simulation study  and the result of the same is reported in Section 4.
\section{Test for right censored case}
Next we discuss how the censored observations can be incorporated in  the proposed testing procedure.
Consider the right-censored  data  $(Y, \delta)$, with $Y=min(X,C)$ and $\delta=I(X\leq C)$, where $C$ is the censoring time. We assume censoring times and lifetimes are independent. Now we need to address the testing problem discussed  in Section 2 based on $n$ independent and identical observations $\{(Y_{i},\delta_i),1\leq i\leq n\}$.  We develop the test using the same  departure measure $\Delta (F)$ given in (\ref{delta}). Consider a symmetric kernel defined by $h(Y_1, Y_2)=\frac{1}{2}(2\max(Y_{1},Y_{2})-2Y_{1}-2Y_{2}+Y_{1}^2+Y_{2}^2)$.  An estimator of  (\ref{delta}) in the right censored case is given by (Datta et al., 2010)
\begin{equation*}\label{delta1c}
\widehat{\Delta}_{c}=\frac{2}{n(n-1)}\sum_{i=1}^{n}\sum_{j<i;j=1}^{n}\frac{h(Y_{1},Y_{2})\delta_i\delta_j}{\widehat{K}_{c}(Y_i)\widehat{K}_{c}(Y_j)},
\end{equation*}
 provided $\widehat{K}_{c}(Y_i)>0$ and $\widehat{K}_{c}(Y_j)>0$, with probability 1 where $\widehat{K}_c$ is the Kaplan-Meier estimator of $K_c$, the survival function of  $C$.
  We reject $H_{0}$ in favour of $H_{1}$ for large values of $|\widehat{\Delta}_c|$.

Next we obtain the limiting distribution of $ \widehat{\Delta}_{c}$. Let $N_i^c(t)=I(Y_i\leq t, \delta_i=0)$ be the counting process corresponds to the censoring variable $C_i$, $R_i(t)=I(Y_i\geq t)$. Also let $\lambda_c$ be the  hazard rate of $C$.  The martingale associated with this counting process $N_i^c(t)$ is given by
\begin{equation*}
M_i^c(t)=N_i^c(t)-\int_{0}^{t} R_i(u) \lambda_c(u) du.
\end{equation*}
Denote  $h_1(x)=E(h(X_1,X_2|X_1=x))$, $G(x,y)=P(X_{1}\leq x, Y_{1}\leq y,  \delta=1), x\in \mathcal{X}$, $\bar H(t)=P(Y_{1}> t)$ and
\begin{equation*}
w(t)=\frac{1}{\bar{H}(t)} \int_{\mathcal{X}\times[0,\infty)}{\frac{h_1(x)}{K_c(y-)}I(y>t)dG(x,y)}.
\end{equation*}

\begin{theorem}\label{thm5.1} If  $E(Y_1^2)<\infty$ and the integrals $\int_{\mathcal{X}\times[0,\infty)}{\frac{h^{2}_1(x)}{K_c^2(y)}dG(x,y)}$ and  $\int_0^\infty w^2(t)\lambda_c(t)dt$ are finite, then as $ n \rightarrow \infty $, $\sqrt{n}(\widehat{\Delta}_c-\Delta)$  converges in distribution to  Gaussian with mean zero and variance $4\sigma_{c}^{2}$, where $\sigma_{c}^2$  is given by
\begin{equation*}
\sigma_{c}^{2}=Var\left(\frac{\Big(2XF(X)+2\int_{X}^{1}ydF(y)-2X+X^2\Big)\delta_1}{4K_c(Y_1-)}+\int w(t) dM_1^c(t)\right).
\end{equation*}
\end{theorem}
\begin{corollary}
  Assume that the condition stated in Theorem 3 holds. Under $H_0$,   as $ n \rightarrow \infty $, the distribution of $\sqrt{n}\widehat{\Delta}$  is Gaussian with mean zero and variance $4\sigma_{c0}^{2}$, where $\sigma_{c0}^2$  is given by
\begin{equation*}
\sigma_{c0}^{2}=Var\Big(\frac{X(X-1)\delta_1}{K_c(Y_1-)}+\int w(t) dM_1^c(t)\Big).
\end{equation*}
\end{corollary}

Next we find an estimator of  $\sigma_{c0}^{2}$ using the reweighed techniques. An estimator of $\sigma_{c0}^2$ is given by
\begin{equation*}\label{ecvar}
\widehat{\sigma}_{c0}^2=\frac{4}{(n-1)}\sum_{i=1}^{n}(V_{i}-\bar V)^2,
\end{equation*}
where
\begin{equation*}\label{36}
V_{i}=\frac{\widehat{h}_1(X_{i})\delta_i}{\widehat{K}_c(Y_{i})}+\widehat w(X_{i})(1-\delta_i)-\sum_{j=1}^{n}\frac{\widehat w(X_{i})I(X_{i}>X_{j})(1-\delta_i)}{\sum_{i=1}^{n}I(X_{i}>X_{j})},
\end{equation*}
$$\bar V =\frac{1}{n}\sum_{i=1}^{n}V_{i}, \quad  \widehat{h}_1(X)=\frac{1}{n}\sum_{i=1}^{n}\frac{h(X,Y_{i})\delta_i}{\widehat{K}_c(Y_{i}-)}, \quad R(t)=\frac{1}{n}\sum_{i=1}^{n}I(Y_{i}>t)$$ and
$$\widehat w (t)=\frac{1}{R(t)}\sum\limits_{i=1}^{n}\frac{\widehat{h}_1(X_{i})
\delta_{i}}{\widehat{K}_c(Y_{i})}I(X_{i}>t).$$

Under right censored situation, we reject the null hypothesis $H_{0}$ against the alternative hypothesis $H_{1}$ at a significance level $\alpha$, if
\begin{equation*}
 \frac{ \sqrt{n} |\widehat{\Delta}_c| }{\widehat{\sigma}_{0c}}>Z_{\alpha/2}.
  \end{equation*}
The results of the Monte Carlo Simulation which asses the finite sample performance of the test  is also reported in Section 4.

\section{Empirical evidence}
To evaluate the finite sample performance of our test procedures, we conduct a Monte Carlo simulation study using R software. The simulation is repeated ten thousand times. For complete data, to show the competitiveness of our test, we compare the empirical power our test with the existing test procedures.
\subsection{Uncensored case}
We find the empirical type I error and power of the proposed tests for different choices of alternatives.
We compare our new test with the classical procedures based on the empirical distribution function Kolmogorov-Smirnov  test and the well used tests for uniform distribution proposed by Sherman (1950), Quesenberry and Miller Jr. (1977) and Frozini et al. (1987). Next we briefly discuss about these tests.  The Kolmogorov-Smirnov test statistic is given by $ KS=\text{max}\{D^{+}, D^{-}\}$ where
\begin{equation*}
  D^{+}=\max_{j=1,2,...,n}\left(\frac{j}{n}-\widehat F(X_{(j)})\right)~~~~\text {and}~~~~
  D^{-}=\max_{j=1,2,...,n}\left (\widehat F(X_{(j)})-\frac{j-1}{n}\right)
\end{equation*}
with $\widehat F(.)$ as the empirical distribution function of the observed data. Frozini's test statistic has the form
\begin{equation*}
    F=\frac{1}{\sqrt n}\sum_{j=1}^{n} \left|X_{(j)}-\frac{j-0.5}{n} \right|.
\end{equation*}
Sherman test statistic is given by
\begin{equation*}
   S=\frac{1}{2}\sum_{j=1}^{n+1} \left|(X_{(j)}-X_{(j-1)}) -\frac{1}{n+1} \right|.
\end{equation*}
with $X_{(0)}=0$ and $X_{(n+1)}=1$. The Q-Statistic proposed by Quesenberry and Miller Jr. (1977) is given by
\begin{equation*}
    Q=\sum_{j=1}^{n+1}\left(X_{(j)}-X_{(j-1)}\right)^{2}+\sum_{j=1}^{n}(X_{(j+1)}-X_{(j)})(X_{(j)}-X_{(j-1)}).
\end{equation*}
\begin{table}[h]
\caption{Empirical type I error of the tests}
\scalebox{0.8}{
\begin{tabular}{cccccccccccccc}\hline {t1}
\multirow{2}{*}{} & \multicolumn{2}{c}{$\widehat\Delta$} & \multicolumn{2}{c}{$KS$}&\multicolumn{2}{c}{$F$}&\multicolumn{2}{c}{$S$}&\multicolumn{2}{c}{$Q$}  \\ \cline{1-11}
$n$ &   1\% level  & 5\% level&   1\% level  & 5\% level & 1\% level  & 5\% level&   1\% level  & 5\% level&1\% level  & 5\% level \\ \hline
25&  0.0108& 0.0546 &0.0113 &0.0486& 0.0108 & 0.0513 & 0.0109 &0.0516 &0.0085 &0.0475\\
50&  0.0106& 0.0534 &0.0108 &0.0491& 0.0094 & 0.0464 & 0.0088 &0.0485 &0.0113 &0.0483\\
75&  0.0104& 0.0492 &0.0096 &0.0509& 0.0104 & 0.0474 & 0.0095 &0.0498 & 0.0108 &0.0511\\
100& 0.0102& 0.0502 &0.0103 &0.0508& 0.0102 & 0.0486 & 0.0095 &0.0502 &0.0096 &0.0504\\ \hline

\end{tabular}}
\end{table}
\begin{table}[h]
\caption{Empirical power the tests: $U(0,1.2)$}
\scalebox{0.8}{
\begin{tabular}{cccccccccccccc}\hline
\multirow{2}{*}{} & \multicolumn{2}{c}{$\widehat\Delta$} & \multicolumn{2}{c}{$KS$}&\multicolumn{2}{c}{$F$}&\multicolumn{2}{c}{$S$}&\multicolumn{2}{c}{$Q$}  \\ \cline{1-11}
$n$ &   1\% level  & 5\% level&   1\% level  & 5\% level & 1\% level  & 5\% level&   1\% level  & 5\% level&1\% level  & 5\% level \\ \hline
25& 0.5335&0.6950 &0.2374 &0.3221&0.2643 &0.4635 &0.6012 & 0.7182& 0.6721  &0.7682&\\
50& 0.8146&0.9068 &0.3282 &0.5927&0.4956 &0.7171 &0.8932 & 0.9321&0.8934&0.9302&\\
75& 0.9285&0.9682 &0.5599 &0.7997&0.6682 &0.8562 &0.9431 & 0.9732& 0.9473&0.9651&\\
100&0.9882&0.9921 &0.7481 &0.9212&0.8036 &0.9387 &0.9865 & 0.9921&0.9832&0.9972&\\ \hline
\end{tabular}}
\end{table}


\begin{table}[h]
\caption{Empirical power the tests: Exp$(1)$}
\scalebox{0.8}{
\begin{tabular}{cccccccccccccc}\hline
\multirow{2}{*}{} & \multicolumn{2}{c}{$\widehat\Delta$} & \multicolumn{2}{c}{$KS$}&\multicolumn{2}{c}{$F$}&\multicolumn{2}{c}{$S$}&\multicolumn{2}{c}{$Q$}  \\ \cline{1-11}
$n$ &   1\% level  & 5\% level&   1\% level  & 5\% level & 1\% level  & 5\% level&   1\% level  & 5\% level&1\% level  & 5\% level \\ \hline
25& 0.9998 &0.9999&0.7744 &0.8925 &0.9939 &0.9981 &0.9982 &1.0000&0.9999&1.0000\\
50& 1.0000 &1.0000&0.9851 &0.9976 &0.9998 &1.0000 &1.0000 &1.0000&1.0000&1.0000\\
75& 1.0000 &1.0000&0.9999 &1.0000 &1.0000 &1.0000 &1.0000 &1.0000&1.0000&1.0000\\
100&1.0000 &1.0000&1.0000 &1.0000 &1.0000 &1.0000 &1.0000 &1.0000&1.0000&1.0000\\ \hline
\end{tabular}}
\end{table}

\begin{table}[h]
\caption{Empirical power the tests: Gamma$(1,2)$}
\scalebox{0.8}{
\begin{tabular}{cccccccccccccc}\hline
\multirow{2}{*}{} & \multicolumn{2}{c}{$\widehat\Delta$} & \multicolumn{2}{c}{$KS$}&\multicolumn{2}{c}{$F$}&\multicolumn{2}{c}{$S$}&\multicolumn{2}{c}{$Q$}  \\ \cline{1-11}
$n$ &   1\% level  & 5\% level&   1\% level  & 5\% level & 1\% level  & 5\% level&   1\% level  & 5\% level&1\% level  & 5\% level \\ \hline
25& 0.9999 &1.0000&0.7841 &0.8981 &0.9994 &0.9988 &0.9999&1.0000&0.9999&1.0000\\
50& 1.0000 &1.0000&0.9863 &0.9974 &1.0000 &1.0000 &1.0000&1.0000&1.0000&1.0000\\
75& 1.0000 &1.0000&0.9997 &0.9999 &1.0000 &1.0000 &1.0000&1.0000&1.0000&1.0000\\
100&1.0000 &1.0000&1.0000 &1.0000 &1.0000 &1.0000 &1.0000&1.0000&1.0000&1.0000\\ \hline
\end{tabular}}
\end{table}

\begin{table}[h]
\caption{Empirical power the tests: Weibull$(1,2)$}
\scalebox{0.8}{
\begin{tabular}{cccccccccccccc}\hline
\multirow{2}{*}{} & \multicolumn{2}{c}{$\widehat\Delta$} & \multicolumn{2}{c}{$KS$}&\multicolumn{2}{c}{$F$}&\multicolumn{2}{c}{$S$}&\multicolumn{2}{c}{$Q$}  \\ \cline{1-11}
$n$ &   1\% level  & 5\% level&   1\% level  & 5\% level & 1\% level  & 5\% level&   1\% level  & 5\% level&1\% level  & 5\% level \\ \hline
25 &0.9999 &1.0000 &0.7789 &0.8982 &0.9943 &0.9983 &0.9999 &1.0000&0.9999&1.0000\\
50 &1.0000 &1.0000 &0.9846 &0.9980 &1.0000 &1.0000 &1.0000 &1.0000&1.0000&1.0000\\
75 & 1.0000&1.0000 &0.9999 &1.0000 &1.0000 &1.0000 &1.0000 &1.0000&1.0000&1.0000\\
100&1.0000 &1.0000 &1.0000 &1.0000 &1.0000 &1.0000 &1.0000 &1.0000&1.0000&1.0000\\ \hline
\end{tabular}}
\end{table}


\begin{table}[h]
\caption{Empirical power the tests: Pareto$(1,1)$}
\scalebox{0.8}{
\begin{tabular}{cccccccccccccc}\hline
\multirow{2}{*}{} & \multicolumn{2}{c}{$\widehat\Delta$} & \multicolumn{2}{c}{$KS$}&\multicolumn{2}{c}{$F$}&\multicolumn{2}{c}{$S$}&\multicolumn{2}{c}{$Q$}  \\ \cline{1-11}
$n$ &   1\% level  & 5\% level&   1\% level  & 5\% level & 1\% level  & 5\% level&   1\% level  & 5\% level&1\% level  & 5\% level \\ \hline
25 &1.0000 &1.0000 &1.0000 &1.0000 &1.0000 &1.0000 &1.0000 &1.0000&1.0000&1.0000\\
50 &1.0000 &1.0000 &1.0000 &1.0000 &1.0000 &1.0000 &1.0000 &1.0000&1.0000&1.0000\\
75 & 1.0000&1.0000 &1.0000 &1.0000 &1.0000 &1.0000 &1.0000 &1.0000&1.0000&1.0000\\
100&1.0000 &1.0000 &1.0000 &1.0000 &1.0000 &1.0000 &1.0000 &1.0000&1.0000&1.0000\\ \hline
\end{tabular}}
\end{table}

For finding the empirical type I error, we generated samples of sizes $n=25,50, 75$ and $100$ from $U(0,1)$.  Empirical type I error of the proposed test and above given tests  for uniform distribution is given in Table 1.  We consider different choices of alternatives for finding empirical  power of the tests and the results of the simulation study are given in Tables 2-6.

The comparative study shows the competitiveness of the newly proposed test to classical procedures. Type I error of the test attains the nominal level and the test shows very good power. For the choice of alternative distribution $U(0,1.2)$, the test has good power which shows that our test captures even a slight deviation from the null distribution. Also, even for small sample size $n=25$, we obtain high power which reaches to unity by $n=50$ for other choices of distributions.

%
%
%


\subsection{Censored case}
The performance of proposed test procedures incorporating right censored observations is also validated through a Mote Carlo simulation study.  We consider a mild censoring situation where 20\% of lifetimes are censored and heavy censored situation where 40\% of the lifetimes are censored. In  censored case, we choose sample sizes $n=50,75,100$ and $200$ and simulation is repeated ten thousand times.  

For finding empirical type I error we simulate observations from $U(0,1)$.  Various choices of alternative are consider for finding empirical power.   In both these cases, we generate censored observations from $U(0,c)$ where $c$ is chosen such way that $P(X>C)=0.2(0.4)$.
Results of the simulation study  with right censored observations are given in Tables 7 and 8. Type I error reaches the chosen significance level in both mild and heavy censored situations.  As in case for complete data, the test statistic points out even a small deviation ($U(0,1.2)$) from the null hypothesis. Power of the test reaches one for other choices of alternative  distributions   even for sample size $n=50$.


\begin{table}[h]
\caption{Empirical type I error and power the test with 20\% censoring}
\scalebox{0.8}{
\begin{tabular}{cccccccccccccc}\hline
\multirow{2}{*}{} & \multicolumn{2}{c}{$U(0,1)$} & \multicolumn{2}{c}{$U(0,1.2)$}&\multicolumn{2}{c}{$Exp(1)$}&\multicolumn{2}{c}{$Weibull(1,2)$}&\multicolumn{2}{c}{$Pareto(1,1)$}  \\ \cline{1-11}
$n$ &   1\% level  & 5\% level&   1\% level  & 5\% level & 1\% level  & 5\% level&   1\% level  & 5\% level&1\% level  & 5\% level \\ \hline
50& 0.0088 &0.0474&0.6993 &0.8303 &0.9999 &1.0000 &1.0000&1.0000&1.0000&1.0000\\
75& 0.0096 &0.0506&0.8635 &0.9387 &1.0000 &1.0000 &1.0000&1.0000&1.0000&1.0000\\
100&0.0106 &0.0489 &0.9397 &0.9761 &1.0000 &1.0000 &1.0000&1.0000&1.0000&1.0000\\
200&0.0106 &0.0499&0.9984 &0.9995 &1.0000 &1.0000 &1.0000&1.0000&1.0000&1.0000\\ \hline
\end{tabular}}
\end{table}


\begin{table}[h]
\caption{Empirical type I error and power the test for 40\% censoring }
\scalebox{0.8}{
\begin{tabular}{cccccccccccccc}\hline
\multirow{2}{*}{} & \multicolumn{2}{c}{$U(0,1)$} & \multicolumn{2}{c}{$U(0,1.2)$}&\multicolumn{2}{c}{$Exp(1)$}&\multicolumn{2}{c}{$Weibull(1,2)$}&\multicolumn{2}{c}{$Pareto(1,1)$}  \\ \cline{1-11}
$n$ &   1\% level  & 5\% level&   1\% level  & 5\% level & 1\% level  & 5\% level&   1\% level  & 5\% level&1\% level  & 5\% level \\ \hline

50& 0.0079 & 0.0469&0.5174 &0.6892 &0.9771 &0.9856 &1.0000&1.0000&1.0000&1.0000\\
75& 0.0091 &0.0488&0.6537 &0.8112 &0.9972 &0.9982 &1.0000&1.0000&1.0000&1.0000\\
100&0.0109 &0.0508&0.8025 &0.9209 &0.9996 &0.9998 &1.0000&1.0000&1.0000&1.0000\\
200&0.0103 &0.0502&0.9839 &0.9963 &1.0000 &1.0000 &1.0000&1.0000&1.0000&1.0000\\ \hline
\end{tabular}}
\end{table}

\section{Data analysis}
\subsection{Uncensored case}
The newly proposed test procedures are applied to a real data set for illustration. We consider the data set discussed in Illowsky and Dean (2018) in Page 317, Table 5.1. The data set consist of smiling times of 55 babies measured in seconds. The data originally follows a uniform distribution $U(0,23)$. Using the proposed procedure, we obtain $\hat\Delta$ as 8157.285. The test static value lies in critical region and we reject the hypothesis that the data follows $U(0,1)$.
Next, to ensure the validity of the proposed procedure, we standardize  the same data to $U(0,1)$. For this transformed data  the value of the test statistic  is obtained as $\hat\Delta = -1.2523$, which belongs to the acceptance region. Hence we accept the null hypothesis that the data follows $U(0,1)$.
\subsection{Censored case}
We consider a real data set on life times of 10 pieces of equipment installed in a system.  We consider the data  given in Table 1.2 of Page 4 in Lawless (2011). Out of the 10 observed lifetimes 3 are randomly right censored.
We analysed the data using the  procedures developed for censored data. The value of test statistic is obtained as $\hat\Delta_{c} =9858.522$, which clearly indicates that the data does not follow uniform distribution.
We then standardize  the same data to $U(0,1)$. For the transformed data we obtain  $\hat\Delta_{c} = -0.3451$ and we accept the null hypothesis that the data comes from $U(0,1)$.

\section{Concluding Remarks}
Using the recently introduced Stein's characterization, a  simple non-parametric  test based on U-statistic theory is developed for testing uniform distribution.  The  test is distribution free. We study the asymptotic properties of the test statistic.   Even though  several tests are available in literature to test the uniformity, none of these tests  incorporated censored samples. We discussed how to incorporate censored data in our methodology. An extensive Monte Carlo simulation study is carried to validate the finite sample performance of the tests procedures. Power comparisons show that the performance of our test is competent with the existing tests in the complete case. Also  the test has well controlled error rate even for small sample sizes. In censored case, even with high percentage of censored data (40\%) our test performs well in terms of empirical power and attains the  size of the test.

\end{document}